\documentclass[aps,prb,twocolumn,superscriptaddress,showpacs, showkeys]{revtex4-1}
\usepackage{graphicx}
\usepackage{amsmath}
\usepackage{multirow}
\usepackage{hyperref}
\begin{document}

%\preprint{}

\title{Influence of Ti doping on the incommensurate charge density wave in 1$T$-TaS$_2$}


%\author{X.M. Chen, P. Abbamonte, Wei Ku, Wei's former student, and others }
%\thanks{}
%\affiliation{UIUC}
\author{X. M. Chen}
\affiliation{\small{Department of Physics and Frederick Seitz Materials Research Laboratory, University of Illinois, 104 South Goodwin Avenue, Urbana, IL 61801, USA}}
\author{A. J. Miller}
\affiliation{\small{Department of Physics and Frederick Seitz Materials Research Laboratory, University of Illinois, 104 South Goodwin Avenue, Urbana, IL 61801, USA}}
\author{C. Nugroho}
\affiliation{\small{Department of Physics and Frederick Seitz Materials Research Laboratory, University of Illinois, 104 South Goodwin Avenue, Urbana, IL 61801, USA}}
\author{G. A. de la Pe{\~n}a}
\affiliation{\small{Department of Physics and Frederick Seitz Materials Research Laboratory, University of Illinois, 104 South Goodwin Avenue, Urbana, IL 61801, USA}}
\author{Y. I. Joe}
\affiliation{\small{Department of Physics and Frederick Seitz Materials Research Laboratory, University of Illinois, 104 South Goodwin Avenue, Urbana, IL 61801, USA}}
\author{A. Kogar}
\affiliation{\small{Department of Physics and Frederick Seitz Materials Research Laboratory, University of Illinois, 104 South Goodwin Avenue, Urbana, IL 61801, USA}}
\author{J. D. Brock}
\affiliation{\small{Department of Applied and Engineering Physics, Cornell University, Ithaca, NY, 14853, USA}}
\author{J. Geck}
\affiliation{\small{
 Synchrotron Studies of Quantum Matter, Leibniz Institute for Solid State and Materials Research, Dresden, Germany}}
\author{G. J. MacDougall}
\affiliation{\small{Department of Physics and Frederick Seitz Materials Research Laboratory, University of Illinois, 104 South Goodwin Avenue, Urbana, IL 61801, USA}} 
\author{S. L. Cooper}
\affiliation{\small{Department of Physics and Frederick Seitz Materials Research Laboratory, University of Illinois, 104 South Goodwin Avenue, Urbana, IL 61801, USA}} 
\author{E. Fradkin}
\affiliation{\small{Department of Physics and Frederick Seitz Materials Research Laboratory, University of Illinois, 104 South Goodwin Avenue, Urbana, IL 61801, USA}} 
\author{D. J. Van Harlingen}
\affiliation{\small{Department of Physics and Frederick Seitz Materials Research Laboratory, University of Illinois, 104 South Goodwin Avenue, Urbana, IL 61801, USA}}
\author{P. Abbamonte}
\affiliation{\small{Department of Physics and Frederick Seitz Materials Research Laboratory, University of Illinois, 104 South Goodwin Avenue, Urbana, IL 61801, USA}}

\date{\today}

\begin{abstract}
We report temperature-dependent transport and x-ray diffraction measurements of the influence of Ti hole doping on the charge density wave (CDW) in 1$T$-Ta$_{1-x}$Ti$_x$S$_2$. Confirming past studies, we find that even trace impurities eliminate the low-temperature commensurate (C) phase in this system. Surprisingly, the magnitude of the in-plane component of the CDW wave vector in the nearly commensurate (NC) phase does not change significantly with Ti concentration, as might be expected from a changing Fermi surface volume. Instead, the angle of the CDW in the basal plane rotates, from 11.9$^\circ$ at $x=0$ to 16.4$^\circ$ at $x=0.12$. Ti substitution also leads to an extended region of coexistence between incommensurate (IC) and NC phases, indicating heterogeneous nucleation near the transition. Finally, we explain a resistive anomaly originally observed by DiSalvo [F. J. DiSalvo, et al., Phys. Rev. B {\bf 12}, 2220 (1975)] as arising from pinning of the CDW on the crystal lattice. Our study highlights the importance of commensuration effects in the NC phase, particularly at $x\sim0.08$.
\end{abstract}

\pacs{71.27.+a}
%\keywords{?}

\maketitle


\section{Introduction}

First studied over forty years ago, the transition metal dichalcogenides (TMDs) have regained widespread attention because of parallels between many of their properties and phenomena observed in high temperature superconducting copper-oxide and iron-arsenide materials\cite{castroneto2001,rossnagel2010,borisenko2008,klemm2000}. In addition to having a layered, quasi-2D structure, these materials exhibit competition between superconductivity (SC) and charge density wave (CDW) order\cite{borisenko2008}, interplay between electron-electron and electron-phonon interactions\cite{rossnagel2010}, ``strange metal" normal state properties, pseudogap effects \cite{borisenko2009,klemm2000}, etc. Of particular interest is the response of such materials to impurity dopants or pressure, which are known to tune the structural and electronic properties of TMD materials \cite{wu1990,wu1988,bando2000,su2002}. Prominent examples include the stabilization of superconductivity by Cu intercalation in 1$T$-TiSe$_2$\cite{morosan2006} and 1$T$-TaS$_2$\cite{wagner2008}, as well as pressure induced superconductivity and its coexistence with CDW in Ti doped 1$T$-TaS$_2$ \cite{sipos2008,ritschel2013}. 

\begin{figure}
\center{\includegraphics[width=0.5\linewidth,angle=-90,clip]{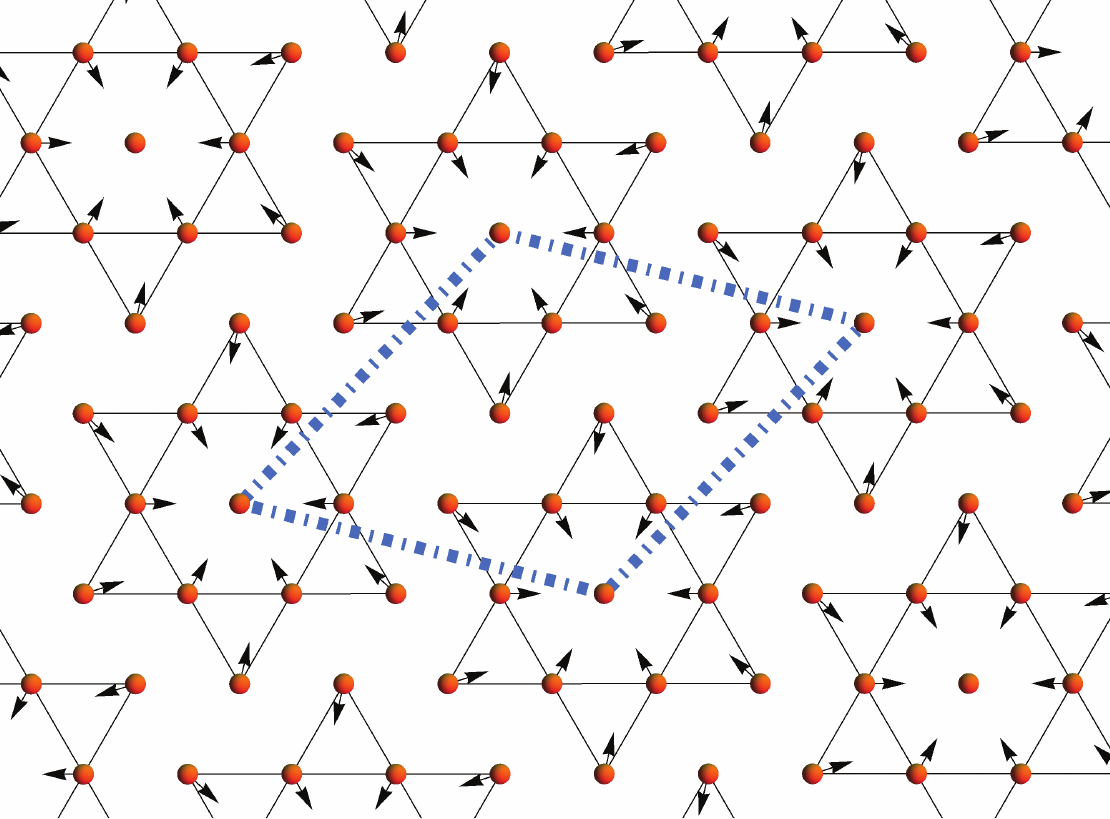}}
\caption[]{Pattern of CDW displacements in the Ta layer of 1$T$-TaS$_2$ in the low-temperature commensurate phase (orange circles represent Ta atoms). The blue dashed line represents a repeating unit of the superstructure. The incommensurate phase studied in this work can be thought of as arising from ordered defects in this commensurate structure.\cite{spijkerman1997}
} \label{david}
\end{figure}

Here we present a x-ray structural study of the effects of Ti doping on the CDW in 1$T$-TaS$_2$. TaS$_2$ exhibits the most intricate CDW phenomenology of all the dichalcogenides, having three distinct phases including an incommensurate (IC) phase that sets in below $T_{IC}$=1120 K, a nearly commensurate (NC) phase below $T_{NC}$=350 K, and finally a commensurate (C) phase below $T_C$=180 K \cite{rossnagel2010,thomson1988}. The lowest transition is first order and is usually attributed to a Mott transition in which the remaining valence electrons (that have not already been gapped by the NC phase) crystallize into a static lattice with superlattice parameter $\sqrt{13} \, a_o$\cite{fazekas1979,rossnagel2010,sipos2008}. This phase is characterized by a significant Peierls distortion consisting of repeating units with the structure of a Star of David (Fig. \ref{david}), and might also be thought of as a polaron lattice. The nominal CDW wave vectors of these phases are summarized in Table \ref{pureTaS2}.

\begin{table}[b]
\begin{tabular}{l | c c c}
 CDW \ \     &\ \ \ \ CDW wave vector ${\bf q}_{||}$ \ \ \ \  & \ \ \ \ $q_{||}$ \ \ \ \ & \ \ \ $\phi$\ \ \  \\
 \hline
 IC               & $0.283$\textbf{$a_o^*$} & 0.283 & 0   \\
 NC            & $0.245$\textbf{$a_o^*$} + $0.068$\textbf{$b_o^*$} & 0.2816 & $\sim$12 \\
 C           & $\frac{3}{13}$\textbf{$a_o^*$} + $\frac{1}{13}$ \textbf{$b_o^*$} & 0.2774 & $\sim$13.9  \\

 \end{tabular}
     \caption{Nominal in-plane component of the CDW wave vector in undoped TaS$_2$. $q_{||}$ represents the magnitude and $\phi$ the azimuthal angle in the basal plane (see Fig. \ref{6CDWfig}). \textbf{$a_o^*$}, \textbf{$b_o^*$} and \textbf{$c_o^*$} are reciprocal lattice vectors associated with real space basis vectors $a_o$, $b_o$ and $c_o$. The wave vector in the third direction stays very close to $c_0^*/3$. }
   \label{pureTaS2}
\end{table}

%\begin{figure}
%\center{\includegraphics[width=0.5\linewidth,angle=-90,clip]{structure.pdf}}
%\caption[]{The basic structure of 1T-TaS2. 1T-TaS2 is trigonal,
%space group $P3_m1$, taken from Spijkerman and Thompson.} \label{structure}
%\end{figure}

The influence of Ti doping was originally studied by DiSalvo \cite{disalvo1975,wilson1975}, where it was shown that Ti substitutes for Ta, hole-doping the system. They also showed that TaS$_2$ is unstable with respect to the formation of S vacancies, which electron dope the system. Hence, unless steps are taken to keep the S content stoichiometric, through the use of precisely controlled S overpressure\cite{disalvo1973}, the composition should be written Ta$_{1-x}$Ti$_x$S$_{2-\delta}$, where $\delta$ is a small number. Even trace quantities of either Ti impurities or S vacancies eliminate the C phase altogether\cite{disalvo1975,zwick1998}. This may be because the additional carriers screen the Coulomb interactions responsible for Mott localization\cite{fazekas1979}. Ti doping also reduces $T_{NC}$ and broadens this transition. 

One intriguing observation by DiSalvo \cite{disalvo1975} is that the resistance change at $T_{NC}$, characterized by the ratio $\rho(4.2 K)/\rho(360 K)$, exhibits a peak at $x=0.08$. The enthalpy change at the transition, $\Delta H_{NC}$, also exhibits a peak at a similar composition\cite{disalvo1975}. These effects been attributed to cation ordering\cite{fazekas1979}, which might exhibit a period of $a=\sqrt{0.08}\,a_0 \sim a_0/\sqrt{13}$ and would stabilize a CDW of the same wave vector. However, no evidence for structural ordering of cations, which would appear as a diffuse ring of scattering in x-ray experiments, has been observed in this system. 

To understand the origin of the resistance peak, and the effects of Ti doping on TaS$_2$ more generally, it is crucial to measure the composition dependence of the wave vector of the CDW. Ti substitution should decrease the Luttinger volume of the Fermi surface and, presumably, the size of whatever nesting vectors may be relevant to CDW formation (if any). Wave vector studies would, then, provide insight into changes in Fermi surface topology, and shed light on the origin of CDW formation in pure TaS$_2$ as well as doped materials.

\section{Experiment}
Single crystals of Ta$_{1-x}$Ti$_x$S$_{2-\delta}$ with $x=0$, 0.04, 0.08, and 0.12 were grown using iodine vapor transport techniques described previously by the same growers.\cite{disalvo1975} No special methods were used to fix the S stoichiometry, so we expect these crystals to contain trace amounts of S vacancies.

Prior to x-ray studies, the resistivity of the crystals was characterized with 4-terminal transport measurements. The crystals were prepared by attaching the as-grown flakes onto a glass slide and making contacts with conductive silver paint in a Hall bar geometry. Indium was used to contact leads to the pads. The samples were current-biased and the differential voltage was read-out using the standard lock-in technique. The samples were cooled to $T=4.2 K$ in a liquid helium flow cryostat and the temperature swept using resistive heater. 

\begin{figure}
\center{\includegraphics[width=1.0\linewidth,angle=0,clip]{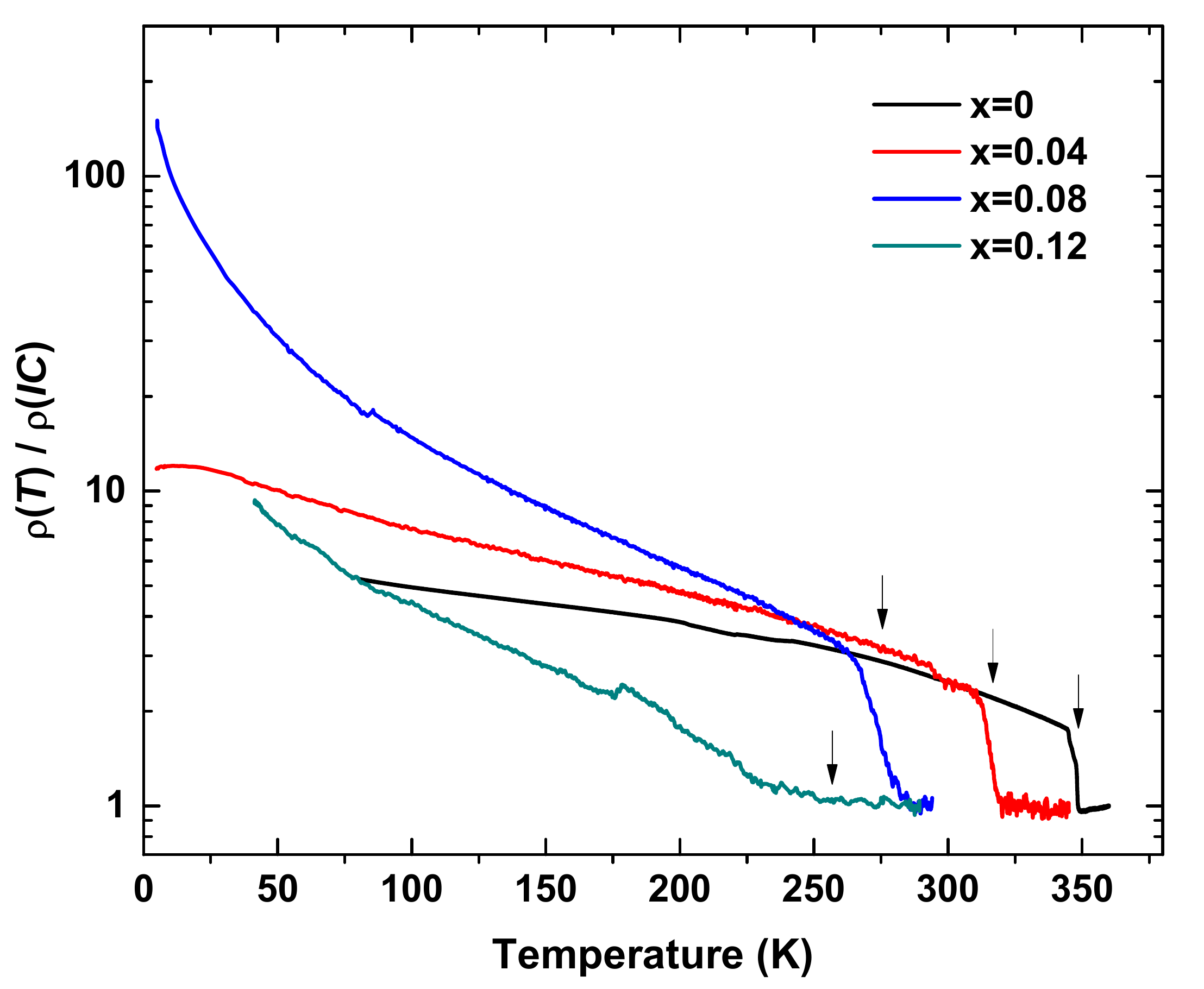}}
\caption[]{Temperature dependence of the resistivity, $\rho(T)$, of Ta$_{1-x}$Ti$_x$S$_{2-\delta}$ for the four compositions studied. Curves are normalized to the value in the IC phase, $\rho(IC)$. The $T_{NC}$ values for each curve are indicated with black arrows. $T_{NC}$
is reduced monotonically by Ti doping, though the size of the resistance jump is largest at $x=0.08$ (see Fig. \ref{peak8}). Note that the commensurate transition at $T_C$ is not observed in these materials (see text).
} \label{resistivity}
\end{figure}

%FOR FUTURE REFERENCE: Here are the T_NC values for the four compositions. Should we state these in the text?
%T0=347.5
%T4=315.6
%T8=273.2
%T12=260 

\begin{figure}
\center{\includegraphics[width=1.0\linewidth,angle=0,clip]{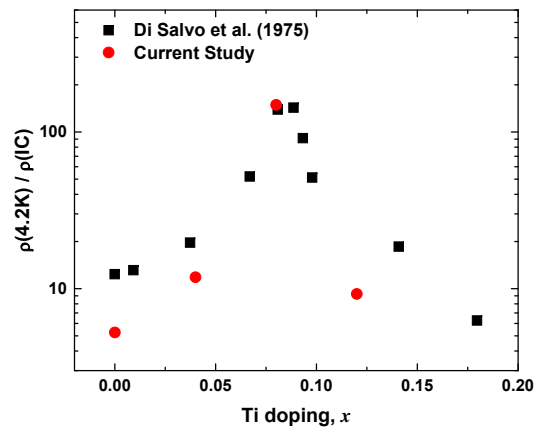}}
\caption[]{Magnitude of the resistance jump at $T_{NC}$, characterized by the ratio $\rho(4.2 K)/\rho(360 K)$, as a function of Ti concentration, $x$. (black squares) Early results reported by DiSalvo [Ref. \cite{disalvo1975}]. (red circles) Current study. The ratio exhibits a maximum at $x\sim 0.08$.} \label{peak8}
\end{figure}

Confirming the results of previous studies,\cite{disalvo1975} Ti doping was found to reduce $T_{NC}$ and to broaden this transition, as illustrated in Fig. \ref{resistivity}. Moreover, no C phase was observed, which suggests that our samples contain a small but unknown concentration of S vacancies. Nevertheless, the intriguing peak in the resistance ratio, $\rho(4.2 K)/\rho(360 K)$, is reproduced in our data at the same value $x=0.08$ reported previously\cite{disalvo1975}, as shown in Fig. \ref{peak8}. Therefore, it is still possible with our materials to investigate this particular phenomenon. 

X-ray experiments were done with MoK$_\alpha$ radiation (17.4 keV) from an 18 kW Rigaku RU-300 rotating anode source. Samples were cooled with a closed-cycle cryostat mounted to a Huber four-circle diffractometer. The angular resolution of the instrument was not sufficient to investigate lineshape changes in the CDW, but was adequate for measuring integrated intensities, which are the focus of the current study. The high temperature symmetric structure of 1$T$-TaS$_2$ consists of identical three-atom-thick TaS$_2$ sheets that are bound by Van der Waals forces. Each layer has a trigonal antiprismatic structure, but overall octahedral symmetry over two layers, with lattice parameters $a_o$ =$b_o$ = 3.36 {\AA} and $c_o$ = 5.90 {\AA}.

\begin{figure}
\center{\includegraphics[width=1.0\linewidth,angle=0,clip]{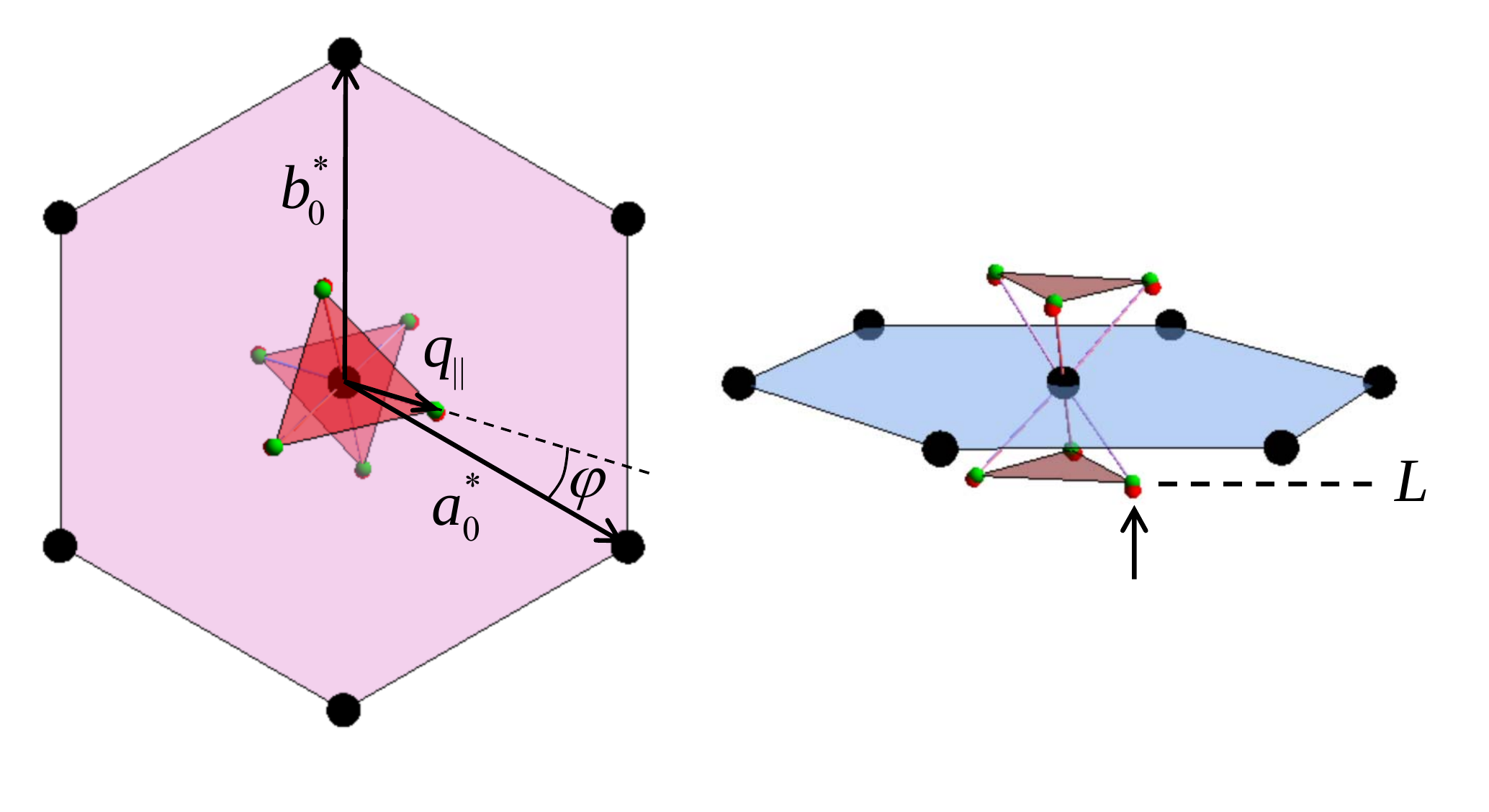}}
\caption[]{Illustration of the momentum space location of the CDW reflections in the vicinity of the (0,0,3) crystalline Bragg peak, with the (3/13, 1/13, 8/3) reflection examined in the current study indicated with an arrow. We describe the wave vector of this reflection in terms of the in-plane magnitude, $q_{||}$, the azimuthal angle in the basal plane, $\phi$, and the periodicity along the $c$ axis, $L$, as shown.} \label{6CDWfig}
\end{figure}

The signature of a CDW in x-ray experiments is the appearance of a superlattice reflection below the ordering transition whose momentum appears at fractional Miller indices, i.e., ${\bf Q} = H a_0^* + K b_0^* + L c_0^*$, where either $H$, $K$, or $L$ is not an integer, and $a_0^*$, $b_0^*$ and $c_0^*$ are Bravais vectors of the reciprocal lattice. Following previous conventions \cite{spijkerman1997, scruby1975}, we will express ${\bf Q}$ in terms of the magnitude of the in-plane wave vector, $q_{||}$, the angle of this vector in the basal plane, $\phi$, and the third Miller index, $L$, which describes the stacking periodicity in the $z$ direction. These quantities are illustrated in Fig. \ref{6CDWfig}. The nominal wave vectors of the various CDW phases in pure TaS$_2$ are given in Table \ref{pureTaS2}. In the current study we will focus on the reflection that lies closest to the (3/13, 1/13, 8/3) commensurate point, which is one of the satellites of the $(0,0,3)$ structural Bragg peak.

\section{Results}

\begin{figure}
\center{\includegraphics[width=0.9\linewidth,clip]{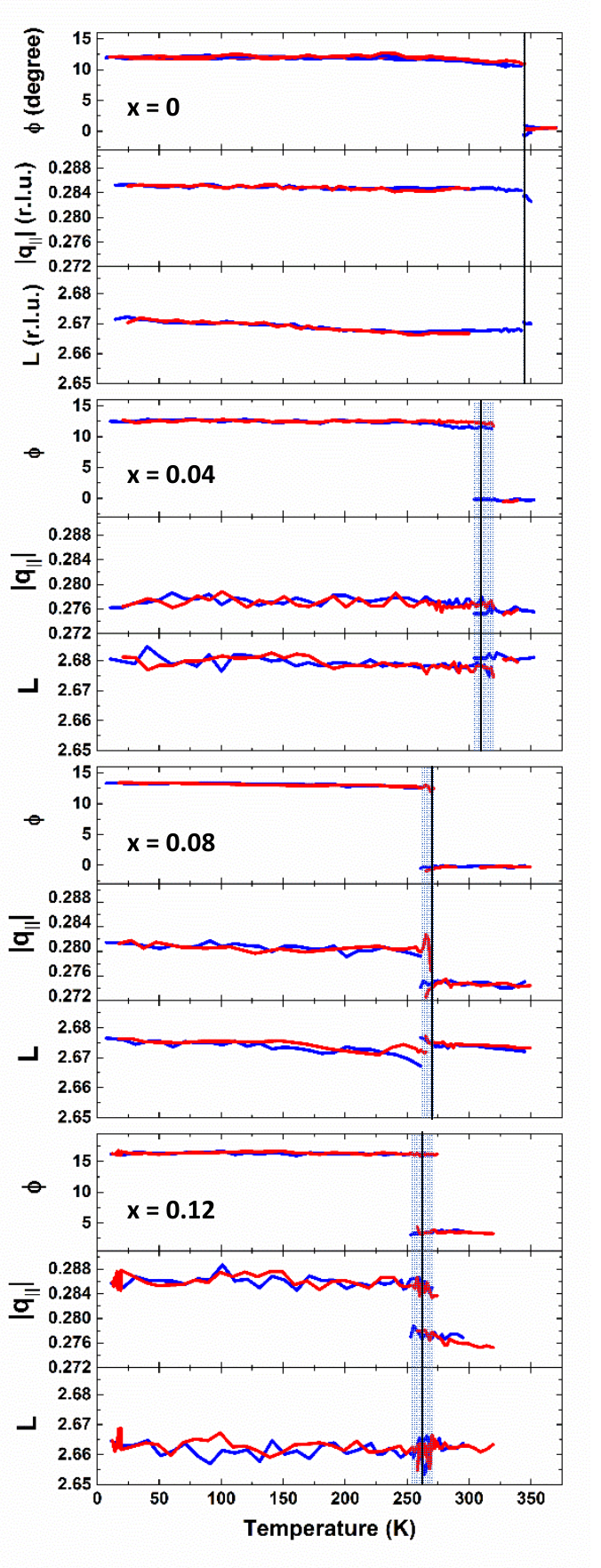}}
\caption[]{Temperature dependence of the CDW wave vector in Ta$_{1-x}$Ti$_x$S$_{2-\delta}$ for all four compositions studied. Red and blue curves represent heating and cooling cycles, respectively. The thin vertical lines indicate the value of $T_{NC}$ for each composition (see Fig. \ref{resistivity}). 
The primary effect of doping is to reduce $T_{NC}$ and broaden the transition, leading to a coexistence region in which both IC and NC phases are present (shaded blue regions).} \label{hkl4}
\end{figure}

The behavior of the CDW wave vector as a function of temperature for all four samples---both for heating and cooling cycles---is displayed in Fig. \ref{hkl4}. Apart from the absence of the $C$ phase, which does not occur in our materials, results for the TaS$_{2-\delta}$ system are very similar to the past observations of Scruby.\cite{scruby1975} Changes in the magnitude, $q_{||}$, are small across the entire temperature range. The azimuthal angle, on the other hand, abruptly rotates from $\phi=0$ to $\phi=12.4^\circ$ when cooling through $T_{NC}$, and continuously changes as the temperature is lowered. The CDW ``stacking" along the $c$ direction stays close to the commensurate value, $L=8/3=2.67$ (the slight changes visible are due to thermal expansion of the sample stage).

As Ti is added, the azimuthal angle in the IC phase stays fixed at approximately $\phi=0$. In the NC phase, however, $\phi$ changes significantly with Ti concentration, from approximately $11.9^\circ$ at $x=0$ to appoximately $16.4^\circ$ at $x=0.12$. Further, at higher doping levels, the value of $q_{||}$ abruptly changes at $T_{NC}$. This stands in contrast to the undoped case, in which only the $\phi$ angle changes at the transition. 

%The stacking of the CDW along the $c$ axis also is affected by doping. In the $x=0.04$ sample the stacking becomes incommensurate, the CDW peak being observed at $L=2.77$. This suggests the presence of periodic stacking faults, with a period of $\sim 10 c_0$. Strangely, this is only observed in the $x=0.04$ sample; the stacking in the $x=0.08$ and $x=0.12$ samples remains commensurate, with $L=2.67$ (Fig. 5). 

\begin{figure}
\center{\includegraphics[width=0.9\linewidth,clip]{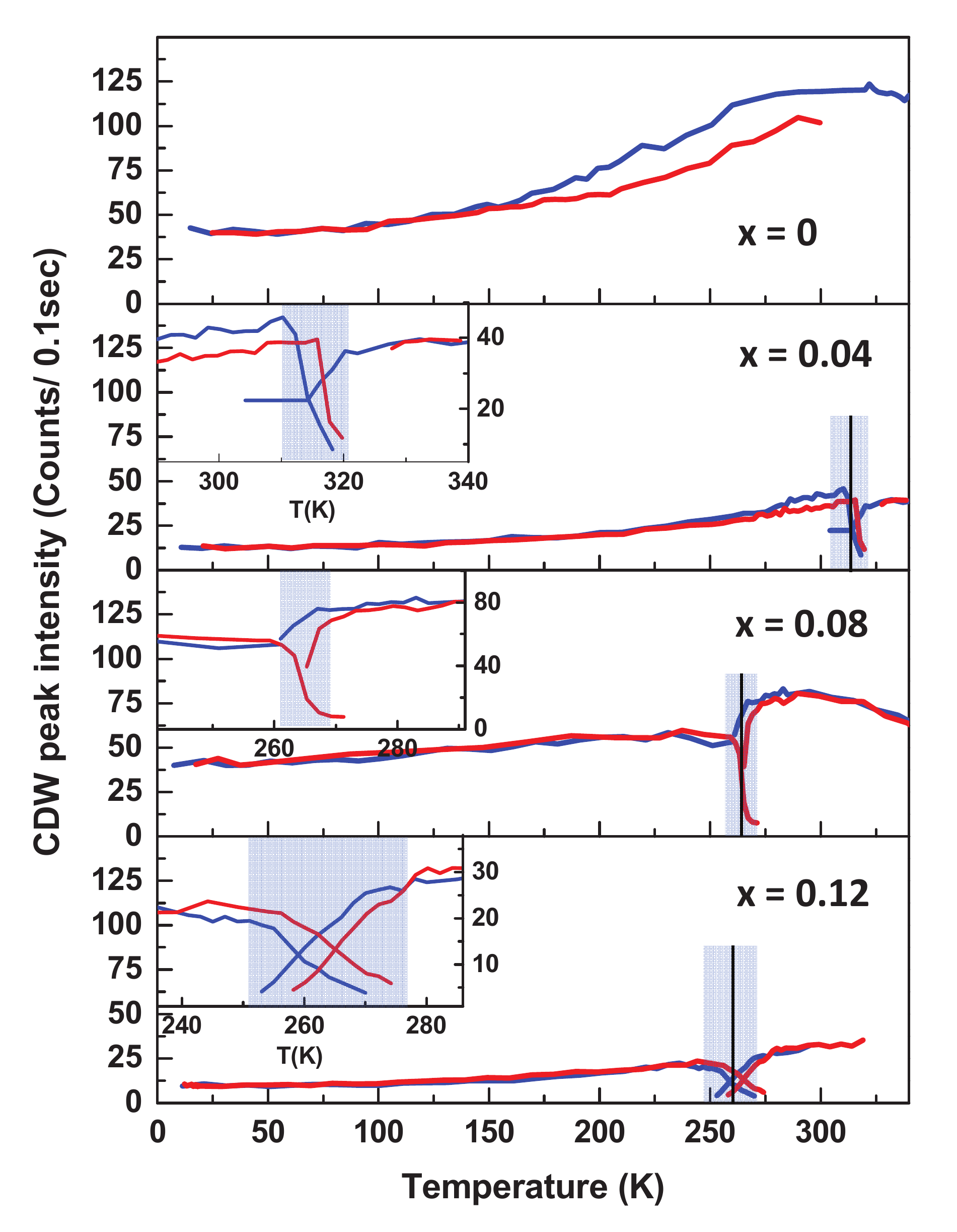}}
\caption[]{Temperature dependence of the intensity of the CDW reflection in Ta$_{1-x}$Ti$_x$S$_{2-\delta}$ for all four compositions studied. Red and blue curves represent heating and cooling cycles, respectively. The thin vertical lines indicate the value of $T_{NC}$ for each composition (see Fig. \ref{resistivity}), and the coexistence regions are shaded in blue. Note that the CDW intensity at $x=0.08$ is greater than that at $x=0.04$ or $x=0.12$, which we attribute to commensuration effects (see text).} \label{cdwPeak}
\end{figure}

The integrated intensity of the CDW reflection, which characterizes the amplitude of the CDW, was also found to be dependent on temperature and doping (Fig. \ref{cdwPeak}). At $x=0$ some hysteresis was observed in the CDW intensity, though not in the value of $T_{NC}$. Like the resistance ratio showed in Fig. \ref{peak8}, the CDW intensity also exhibits a local maximum at $x \sim 0.08$. This indicates that, while $T_{NC}$ is suppressed in this system by Ti disorder, the CDW amplitude is maximum for this composition. 

\begin{figure}
\center{\includegraphics[width=0.8\linewidth,angle=0,clip]{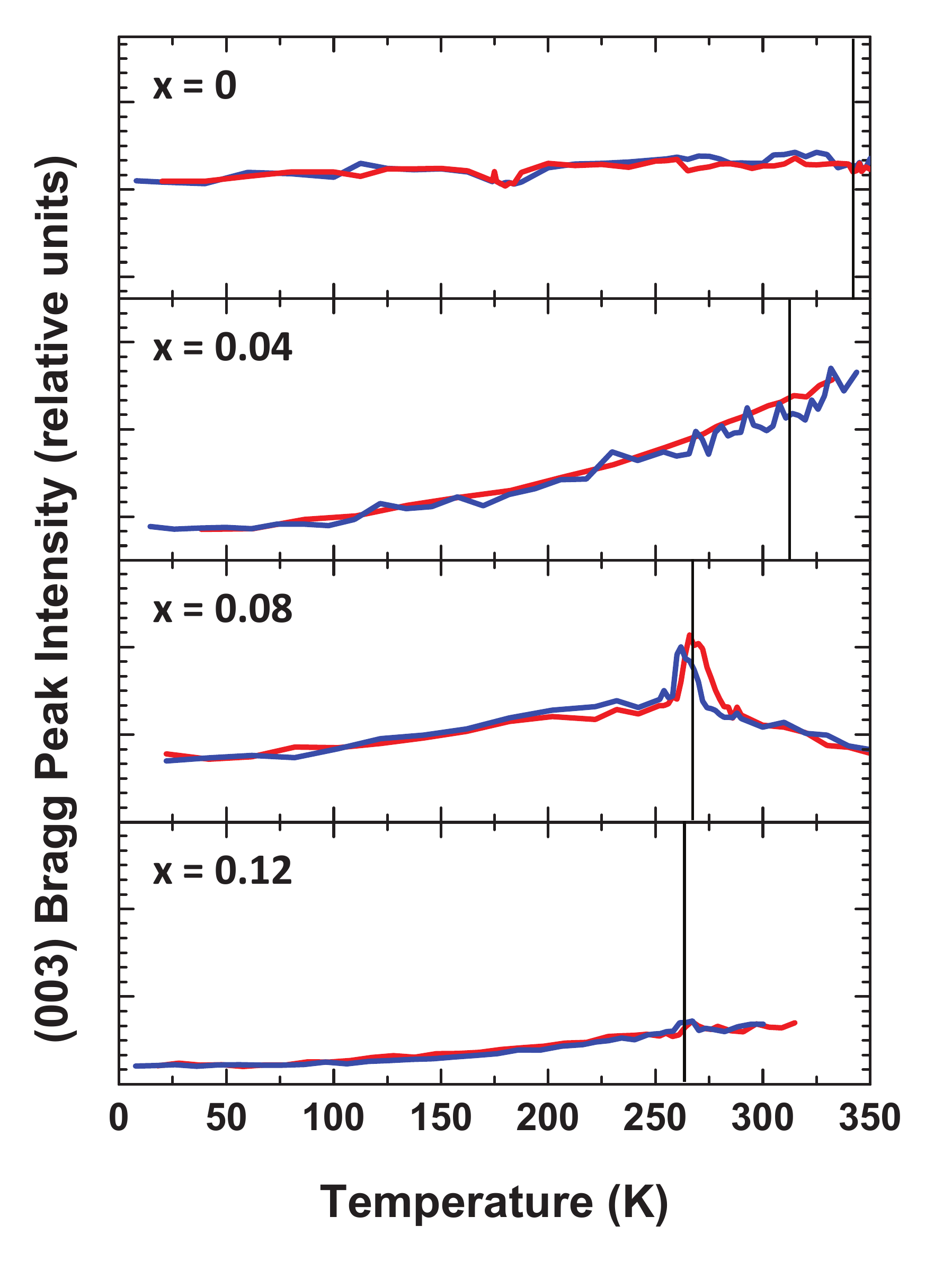}}
\caption[]{Intensity of the (003) structural Bragg peak for different Ti compositions. Red and blue curves represent heating and cooling cycles, respectively. 
The thin vertical lines indicate the value of $T_{NC}$ for each composition (see Fig. \ref{resistivity}).
Note that the crystal with $x=0.08$ exhibits a large anomaly at $T_{NC}$, indicating significant fluctuations in the fundamental lattice near the transition at this composition.} \label{003Peak}
\end{figure}

At higher Ti concentrations, anomalies were also observed in the fundamental, structural lattice at $T_{NC}$. Fig. \ref{003Peak} shows the intensity of the (003) structural Bragg peak as a function of $T$ for all four compositions. At $x=0$ and $0.04$, the temperature dependence is featureless, though at $x=0.04$ the curve is anomalous in that the intensity increases with increasing $T$, which is the opposite of what is expected from a Debye-Waller effect. At higher doping, however, a pronounced intensity anomaly is observed at $T_{NC}$. With its peak intensity ten times stronger than that in $x = 0.12$, the anomaly is largest in the $x=0.08$ sample. This is also the doping at which we have observed the largest CDW amplitude and the largest resistance change at the transition.

%\begin{figure}
%\center{\includegraphics[width=0.65\linewidth,angle=-90,clip]{distFrom003n2.pdf}}
%\center{\includegraphics[width=1.3\linewidth,angle=-90,clip]{distance.pdf}}
%\caption[]{Doping dependence on the length of the of CDW wave vector from (a) their center (003) peak and (b) corresponding C-CDW position in pure TaS$_2$ in both IC and NC phases in r.l.u. The vertical bars indicate one standard deviation. Figure (b) corresponds to $\delta_{q_{//}}$ and $\delta_{\phi}$ combined in Fig \ref{distFromCCDW1}.} \label{distFrom003}
%\end{figure}

%\begin{figure}
%\center{\includegraphics[width=1\linewidth,angle=0,clip]{disalvo.pdf}}
%\caption[]{Previous observation on doped Ti$_x$Ta$_{1-x}$S2 by Di Salvo et al.} \label{disalvo}
%\end{figure}

%\begin{figure}
%\center{\includegraphics[width=0.7\linewidth,angle=-90,clip]{distFromCCDW1.pdf}}
%\caption[]{Doping dependence on the change in $|q_{//}|$ and $\phi$ from IC to NC phase. $\delta_{q_{//}}$ = $|q_{_{IC}}|$ - $|q_{_{NC}}|$, and %$\phi$, $\delta_{\phi}$ = $\phi_{_{NC}}-\phi_{_{IC}}$. The vertical bars indicate one standard deviation. Dotted lines correspond to the calculated %values of distance between IC-CDW and C-CDW. Distance between data points and dotted lines is, therefore, the indication of commensurability.} %\label{distFromCCDW1}
%\end{figure}

\begin{figure}
\center{\includegraphics[width=0.9\linewidth,angle=0,clip]{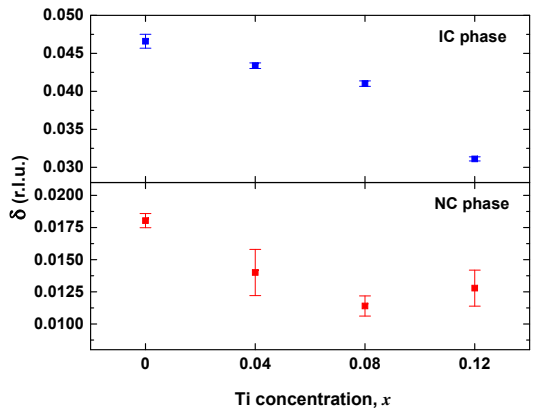}}
\caption[]{Composition dependence of the degree of incommensurability of the CDW, $\delta$, defined as the distance in reciprocal space between the measured wave vector and the commensurate point ${\bf Q}_C = (3/13, 1/13, 8/3)$. (upper panel) Average value in the IC phase. (lower panel) Average value in the NC phase. Note that the CDW is most commensurate for $x=0.08$, where the resistance jump at the transition is maximum.} \label{distFromCCDW2}
\end{figure}

\section{Discussion}

Our results suggest a simple, physical explanation for the origin of the resistive jump at $T_{NC}$, which as shown in Fig. \ref{resistivity} exhibits a maximum at $x \sim 0.08$.\cite{disalvo1975} This effect was speculated by Fazekas\cite{fazekas1979} to be due to pinning on the Ti dopants, which at $x=0.08$ might have a similar periodicity as the CDW. However, no evidence for ordering of the cations, which would appear as a temperature-independent ring of diffuse scattering, has been reported in the literature, nor have we observed it in the current study.

Instead, we refer the reader to Fig. \ref{distFromCCDW2}, which shows the distance, $\delta$, in momentum space between the observed CDW peak position and the nominal commensurate wave vector ${\bf Q}_C = (3/13,1/13,8/3)$. The quantity $\delta$ is a measure of the degree of incommensurability of the CDW, and is given in both the IC and NC phases for the four compositions. The lower panel of Fig. \ref{distFromCCDW2}  (NC phase) shows that $x=0.08$ corresponds to the composition at which $\delta$ is minimum, i.e., at which the CDW is most commensurate. This is also the composition at which the CDW has the largest amplitude (Fig. \ref{cdwPeak}), the largest resistance change at $T_{NC}$ (Fig. \ref{peak8}), and the largest intensity anomaly in the $(0,0,3)$ reflection (Fig. \ref{003Peak}). This anomaly suggests that significant symmetry-breaking fluctuations take place near the transition, at this composition, even in the underlying lattice. Our data suggests that the NC phase is the most stable at $x=0.08$.

We therefore suggest an alternative explanation, which is that the peak shown in Fig. \ref{resistivity} is a commensuration effect. As originally pointed out by McMillan,\cite{mcmillan1977} the free energy of a CDW has a significant contribution from pinning on the lattice, which makes it energetically favorable for a CDW to take on a periodicity that is a commensurate multiple of the undistorted structure. This commensuration energy will increase the energy gap, and lower the overall free energy, of a commensurate state compared to an incommensurate state, and gives rise to the lock-in transitions observed in many dichalcogenides.\cite{rossnagel2010} Our study indicates that the enhanced amplitude and resistance anomaly associated with the CDW at $x=0.08$ arise because the Fermi surface geometry at this composition favors a CDW phase with a wave vector that is close to a commensurate point. As the result, charge mobility decreases at this doping. Note also (Fig. 8) that the degree of commensurability, $\delta$, in the IC phase decreases monotonically Ti content, with no anomaly in $x = 0.08$. This suggests that commensurability plays a less important role in formation of the IC phase than it does in the NC phase.  

Finally, Ti doping was observed to influence the kinetics of the phase transition. In the pure sample, the transition at $T_{NC}$ is narrow in temperature. By contrast, as Ti impurities are added the transition broadens into a crossover region characterized by coexistence of both NC and IC phases (see shaded blue regions in Figs. \ref{hkl4} and \ref{cdwPeak}). We attribute this effect to heterogeneous nucleation driven by nonstoichiometric disorder arising from Ti dopants.

\section{Conclusions}
We have presented temperature-dependent transport and x-ray diffraction measurements of the influence of Ti hole doping on the charge density wave (CDW) in 1$T$-Ta$_{1-x}$Ti$_x$S$_2$ in the nearly commensurate (NC) phase. We observed significant changes in the CDW wave vector as a function of both temperature and Ti doping. Our x-ray scattering results showed a first order phase transition from the IC phase to the NC phase in all doping levels with decreased transition temperature as Ti doping was increased. The doping level $x=0.08$ was found to be special in that it exhibits a peak in the amplitude of the CDW in the NC phase and the largest anomaly in the (0,0,3) structural Bragg reflection, in addition to the large resistance change previously reported.\cite{disalvo1975} We argue that these anomalies arise because the CDW is most commensurate with the lattice at x=0.08.

\begin{acknowledgments}
X-ray experiments were supported by the U.S. Department of Energy, Office of Science, Office of Basic Energy Sciences under Award Number DE-FG02-06ER46285. Transport measurements were supported by award number DE-SC0012368.
\end{acknowledgments}


\begin{thebibliography}{100}
\bibitem{castroneto2001} A. H. Castro Neto, Phys. Rev. Lett. {\bf 86}, 4382 (2001)
\bibitem{klemm2000} R. K. Klemm, Physica C {\bf 341-348}, 839 (2000)
\bibitem{rossnagel2010} K. Rossnagel, J. Phys.: Condens. Matter {\bf 23} 213001 (2011)
\bibitem{borisenko2008} S. V. Borisenko, A. A. Kordyuk, A. N. Yaresko, V. B. Zabolotnyy, D. S. Inosov, R. Schuster, B. Büchner, R. Weber, R. Follath, L. Patthey, and H. Berger, Phys. Rev. Lett. {\bf 100}, 196402 (2008)
\bibitem{borisenko2009} S. V. Borisenko, A. A. Kordyuk, V. B. Zabolotnyy, D. S. Inosov, D. Evtushinsky, B. Büchner, A. N. Yaresko, A. Varykhalov, R. Follath, W. Eberhardt, L. Patthey, and H. Berger, Phys. Rev. Lett. {\bf 102}, 166402 (2009)
\bibitem{morosan2006} E. Morosan, H. W. Zandbergen, B. S. Dennis, J. W. G. Bos, Y. Onose, T. Klimczuk, A. P. Ramirez, N. P. On and R. J. Cava, Nature Phys. {\bf 2}, 544 (2006)
\bibitem{wagner2008} K. E. Wagner, E. Morosan, Y. S. Hor, J. Tao, Y. Zhu, T. Sanders, T. M. McQueen, H. W. Zandbergen, A. J. Williams, D. V. West, and R. J. Cava, Phys. Rev. B {\bf 78}, 104520 (2008)
\bibitem{fazekas1979} P. Fazekas, E. Tosatti, Philosophical Magazine B {\bf 39}, 3 229 (1979)
\bibitem{sipos2008} B. Sipos, A. F. Kusmartseva, A. Akrap, H. Berger, L. Forro, E. Tutis, Nature Mater. {\bf 7}, 960 (2008) 
\bibitem{spijkerman1997} A. Spijkerman, J. L. de Boer, A. Meetsma, G. A. Wiegers, S. van Smaalen, Phys. Rev. B {\bf 56}, 13757 (1997)
\bibitem{disalvo1975} F. J. DiSalvo, J. A. Wilson, B. G. Bagley, J. V. Waszczak, Phys. Rev. B {\bf 12}, 2220 (1975)
\bibitem{disalvo1973} F. J. DiSalvo, B. G. Bagley, J. M. Voorhoeve, J. V. Waszczak, J. Phys. Chem. Solids {\bf 34}, 1357 (1973)
\bibitem{zwick1998} F. Zwick, H. Berger, I. Vobornik, G. Margaritondo, L. Forró, C. Beeli, M. Onellion, G. Panaccione, A. Taleb-Ibrahimi, and M. Grioni, Phys. Rev. Lett. {\bf 81}, 1058 (1998)
\bibitem{scruby1975} C. B. Scruby, P. M. Williams, G. S. Parry, Phil. Mag. {\bf 31}, 255 (1975)
\bibitem{mcmillan1977} W. L. McMillan, Phys. Rev. B {\bf 16}, 643 (1977)
\bibitem{ritschel2013} T. Ritschel, J. Trinckauf, G. Garbarino, M. Hanfland, M. v. Zimmermann, H. Berger, B. Buchner, and J. Geck, Phys. Rev. B {\bf 87}, 125135 (2013)
\bibitem{wu1990} X-L. Wu and C. M. Lieber, Phys. Rev. B {\bf 41}, 1239(R) (1990)
\bibitem{wu1988} X-L. Wu, P. Zhou, and C. M. Lieber, Phys. Rev. Lett. {\bf 61}, 2604 (1988)
\bibitem{bando2000} H. Bando, K. Koizumi, Y. Miyahara and H. Ozaki,  J. Phys. Condens. Matter {\bf 12}, 4353 (2000)
\bibitem{su2002} J-D Su, A. R. Sandy, J. Mohanty, O. G. Shpyrko, and M. Sutton, Phys. Rev. B {\bf 86}, 205105 (2002)
\bibitem{thomson1988} R. E. Thomson, U. Walter, E. Ganz, J. Clarke, A. Zettl, P. Rauch, and F. J. DiSalvo, Phys. Rev. B {\bf 38}, 10734 (1988)
\bibitem{wilson1975} J.A. Wilson, F.J. Di Salvo, S. Mahajan, Advances in Physics {\bf 24}, 2 (1975)



%\bibitem{DiSalvo} F. J. Di Salvo \textit{et al.}, Surface Science 58 (1986) 297-311
%\bibitem{Ishiguro} T. Ishiguro \textit{et al.}, Phys. Rev. B \textbf{44}, 2046 (1991) and Phys. Rev. B \textbf{52}, 759 (1995)
%\bibitem{Wagner} K. E. Wagner \textit{et al.}, Phys. Rev. B \textbf{78}, 104520 (2008)
%\bibitem{Nakanishi} Various references sited as footnotes in each page of this paper.
%\bibitem{Wu} X. L. Wu \textit{et al.}, Phys. Rev. Lett. \textbf{61}, 2604 (1988)

\end{thebibliography}

\end{document}